\documentstyle[epsfig,sprocl,mcite]{article}

\def\lsim{\mathrel{\rlap{\lower2pt\hbox{\hskip1pt\small$\sim$}}
    \raise2pt\hbox{\small$<$}}} 
\def\gsim{\mathrel{\rlap{\lower3pt\hbox{\hskip0pt\small$\sim$}}
    \raise2pt\hbox{\small$>$}}} 

\newcommand{\D}{{\rm d}}

\newcommand{\mev}{{\rm Me}\kern-1.pt{\rm V}}
\newcommand{\gev}{{\rm Ge}\kern-1.pt{\rm V}}
\newcommand{\gevsq}{\mbox{$\mathrm{{\rm Ge}\kern-1.pt{\rm V}}^2$}}

\newcommand{\ppbar}{\mbox{$p\overline{p}$}}
\newcommand{\rhoz}{\mbox{$\rho^0$}}
\newcommand{\jpsi}{\mbox{$J/\psi$}}
\newcommand{\xbj}{\mbox{$x$}}
\newcommand{\qsq}{\mbox{$Q^2$}}

\newcommand{\phih}{\mbox{${\phi}_h$}}
\newcommand{\Phih}{\mbox{${\Phi}_h$}}

\newcommand{\thetah}{\mbox{${\theta}_h$}}

\newcommand{\fsq}{\mbox{$F_2$}}

\newcommand{\abst}{\mbox{$|t|$}}
\begin{document}

\vspace*{-2.5cm}
\hfill {\mbox{BONN-HE-99-04}}\\
\mbox {}\hfill {\mbox{August 1999}}\\*[9mm]
\title{
Results on Diffractive Processes\\ from the HERA Collider Experiments\footnote{Talk
  presented at the EPIC'99 Workshop on Physics with an Electron/Polarized-Ion
  Collider, Indiana University Cyclotron Facility, 8-11 April 1999,
  Bloomington, Indiana, USA}}

\author{James A. Crittenden}

\address{Physikalisches Institut, University of Bonn, Nu{\ss}allee 12, 53115 Bonn, Germany}


\maketitle\abstracts{We review topical results on diffractive processes
from the experiments H1 and  ZEUS at the HERA electron-proton collider.
Emphasis is placed on the phenomenological and 
experimental consequences of the discoveries at HERA for the proposed electron/polarized-ion collider EPIC.}
\section{Introduction}
The technical feasibility and experimental program of an
electron/polarized-ion collider such as the proposed EPIC collider were studied in
detail during the years 1995-1997.\hspace{-0.7ex}~\cite{enc} During these very years, discoveries at the
first electron-proton collider facility ever built, the HERA accelerator
complex at the DESY laboratory in Hamburg, Germany, were establishing a
new field of study in strong-interaction physics: that of
hard diffractive processes.\hspace{-0.7ex}~\cite{hep99_06_518} Thus, these initial considerations of the
physics motivation for a collider such as EPIC could not fully benefit from 
the information provided by measurements at HERA now available, just as was the case for the 
proposed experimental program at the HERA collider 
itself prior to its operation.\hspace{-0.7ex}~\cite{heraworkshop} It is the purpose of this 
report to summarize 
these results from the HERA collider experiments H1 and ZEUS, placing
particular emphasis on lessons learned about the physics impact
of the results, which we now know to provide valuable insight into the
dynamics of the strong interaction, and concerning the experimental
requirements for such measurements. We will see that experiments at 
the proposed EPIC
collider would provide essential information complementary to that
obtained from investigations at
HERA, profiting from the experience at this first electron-proton
collider such as to greatly extend  the sensitivity to a wide variety
of aspects of strong-interaction dynamics inaccessible by other means. 

For the purposes of our discussion, we assume the EPIC machine 
to collide 4~{\gev} electrons
with 40~{\gev} protons,
achieving instantaneous luminosities of at least 
$10^{33}$~cm$^{-2}$~s,\hspace*{-0.7ex}$^{-1}$ which
is approximately two orders of magnitude greater than that presently reached at
HERA. The electron-proton center-of-mass energy, $\sqrt{s}$, will be therefore
25~{\gev}, which is also the kinematic limit of the virtual-photon/proton
center-of-mass energy, $W$. Such high luminosity will permit investigation
of a range in photon virtuality, {\qsq}\hspace*{-1.2ex},\hspace*{1.2ex} extending well into the perturbative
regime. Of particular interest is the recently identified transition region
$0.1 \lsim \qsq \lsim 1~\gevsq$\hspace*{-1.5ex}. Assuming a central detector geometry
with a beam aperture of a few centimeters, located approximately 1~m from
the interaction point in the rear direction (the electron flight direction),
the scattered electron  will be detected in
to the central detector for this range in photon virtuality, obviating the need for special-purpose electron
detectors.
The Lorentz factor of the center-of-mass system
in the laboratory frame at EPIC 
will be 1.6, less severe than the factor 2.7 at HERA. This will
ease the constraints on the very forward 
detectors employed to distinguish exclusive from proton-dissociative diffractive
processes, but will require excellent tracking and calorimetry 
in the rear direction in
order to cover the photon fragmentation region with good resolution. The
rapidity range covered at EPIC will be approximately 3 units less than
the 9 units available 
at HERA, with negative consequences for inclusive studies dependent upon
rapidity gap requirements, especially studies of jet production. 
On the other hand, the EPIC kinematics are ideal for investigations
in the rapidly-growing field of exclusive and semi-exclusive processes,
such as vector-meson photo- and electroproduction.\hspace{-0.7ex}~\cite{stmp_140} Of particular interest
are measurements of the helicity structure of these diffractive processes,
and the ability to polarize the electron and proton beams allow
EPIC to play a unique r\^ole in its elucidation. Thus, the center-of-mass
energy sufficient to reach the diffractive regime, and to produce all of the
vector mesons, along with the high luminosity and the ability to survey
the transition region in photon virtuality ensure a rich physics program
for EPIC. We omit discussion of the additional investigations
of QCD
effects made possible by the acceleration of ion beams, since
they have not yet been addressed by measurements at HERA.\hspace*{-0.7ex}~\cite{strikman,hep99_07_221,hep99_08_230}
\vspace*{-1mm}

\section{Phenomenological Issues}
\label{sec:phenomenology}
\vspace*{-1mm}

The phenomenological success of Regge theory in describing diffractive
interactions is well established.\hspace*{-0.7ex}~\cite{collins} However, the past 25 years
have produced persuasive experimental evidence that the fundamental structure
of the strong interaction is described by the quantum field theoretical
treatment implemented in Quantum Chromodynamics (QCD). A thorough understanding
of the relationship between
these two approaches remains elusive and
its attainment is viewed as one of the primary goals of contemporary
particle-physics research. The H1 and ZEUS collaborations have invested
major efforts in characterizing the kinematic regions of validity of these
two highly disparate views of the strong interaction, and in measuring
the transition between the two regimes.\hspace*{-0.7ex}~\cite{dr_98_192}

Regge theory leads to various scaling laws which sharply distinguish it
from the quantum chromodynamical description. The energy dependence of
forward scattering cross sections is characteristically weak, scaling as a 
power law with a power of 
0.32, which corresponds to the intercept of the
Pomeron trajectory determined in hadronic interactions~\cite{pl_296_227}:
$\alpha (t) = 1.08 + 0.25~t$. \mbox{Gribov}~\cite{jetp_14_478} showed that the observed steeply
forward peaking of the differential cross section $\frac{\D\sigma}{\D t}$
may be expected to increase logarithmically 
with energy as a
consequence of unitarity and the analyticity of the scattering
amplitudes. This latter phenomenon is often referred to as "shrinkage",
and has been verified experimentally for a wide variety of diffractive
interactions, including the soft diffractive photoproduction of vector
mesons.

\begin{figure}[htbp]
\begin{center}
\epsfig{file=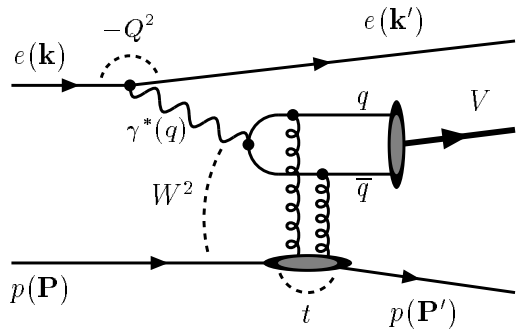, width=6.5cm}
\end{center}
\caption{ 
\label{fig:pqcddiag}
Schematic diagram illustrating exclusive vector-meson electroproduction as
mediated by the
exchange of a gluon pair in a color-singlet state} 
\end{figure}

The past few years have seen rapid progress in the theoretical understanding
of diffractive processes in the framework of QCD, including a number
of impressive phenomenological successes. Diffractive processes are
modeled as proceeding via the exchange of a coherent pair of
gluons in a color-singlet state, as proposed early in the development of
QCD.\hspace*{-0.7ex}~\cite{pr_12_163,*prl_34_1286,*pr_14_246} Figure~\ref{fig:pqcddiag}
shows a diagram for exclusive vector-meson electroproduction which
exemplifies such 
models. Among the many consequences of such a model
are the steepening of the energy dependence of the forward cross section
with increasing photon virtuality due to the gluon density in the target, and 
the energy independence of the forward peaking.\hspace*{-0.7ex}~\cite{shep_11_51} 
Of particular interest 
for the study of exclusive and semi-exclusive meson production
are the specific predictions that the final-state meson be longitudinally
polarized,\hspace*{-0.7ex}~\cite{pr_50_3134,pr_53_3564,*pr_54_5523,pl_449_306,strikman,brodsky} and that small helicity-violating
effects are to be expected.\hspace*{-0.7ex}~\cite{pr_58_114026,schnc_dis99}

Of essential import for these calculations is the identification of 
the scale associated with each type of interaction. The r\^ole of various
kinematic quantities in reaching the asymptotically free regime
characteristic to QCD is under active investigation at HERA. The
photon virtuality,\hspace*{-0.7ex}~\cite{pr_56_2982,hep99_05_226} the vector-meson mass
in exclusive photo- and electroproduction of vector mesons~\cite{zfp_57_89,
  pr_54_3194} and the momentum transferred to the target in exclusive and
semi-exclusive
processes~\cite{pl_65_463,*pr_15_2503,pr_53_3564,*pr_54_5523,pl_449_306} 
have each been proposed as scale variables
The collider experiments at HERA are sensitive to the transition region
in each of these variables, as will be the experiments at the EPIC collider.

\section{Total {$\gamma^* {\rm p}$} cross section}

\begin{figure}[hbp]
\begin{center}
\epsfig{file=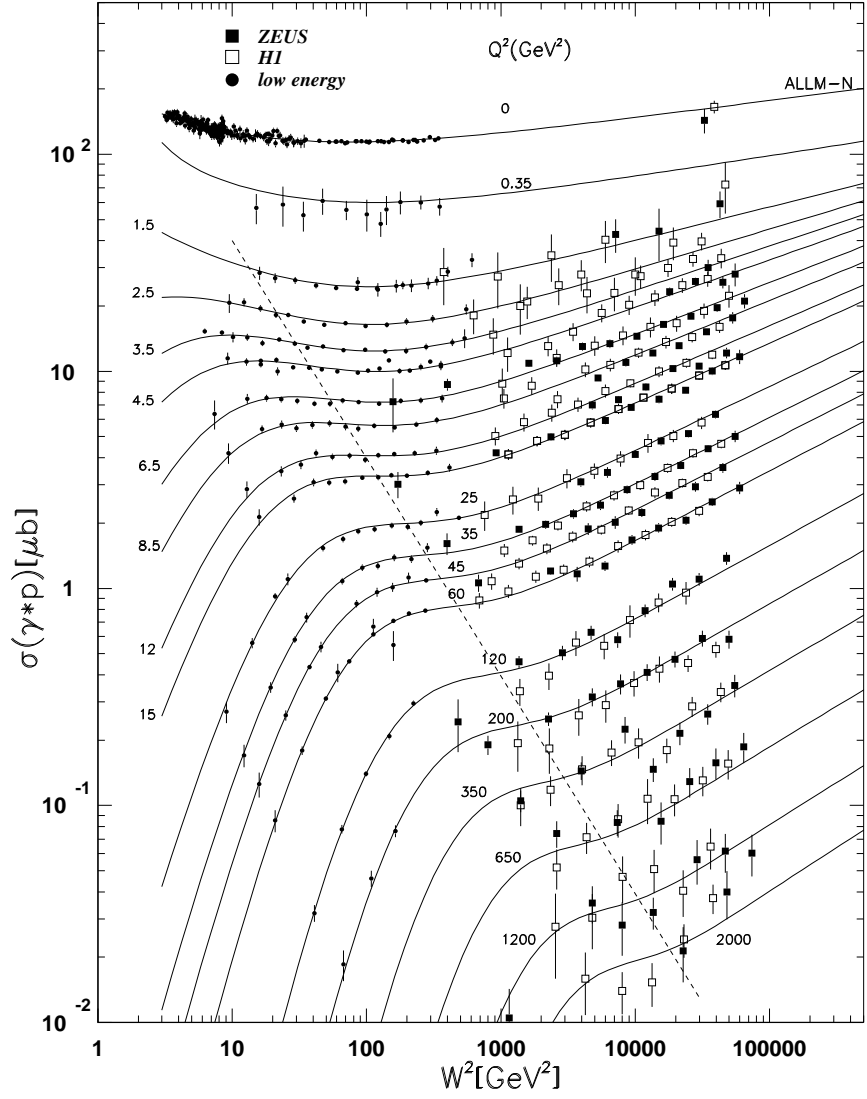, width=0.95\textwidth, bbllx=3, bblly=8, bburx=524, bbury=686, clip=}
\end{center}
\caption{ 
\label{fig:sigtot}
Total photon--proton cross section, $\sigma({{\gamma^*}\rm p})$, as a function of
the squared center-of-mass energy for various values of {\qsq}
as measured by the ZEUS~\protect\cite{zfp_72_399} and H1~\protect\cite{np_470_3} collaborations and by
the NMC collaboration~\protect\cite{pl_364_107} at CERN. The curves represent
calculations using the ALLM proton structure
function par\-a\-me\-tri\-za\-tions.\hspace*{-0.7ex}~\protect\cite{pl_269_465,*marcusthesis}
The dashed line connects points where {\xbj}=0.06
} 
\end{figure}

In order to study the transition from the soft to the hard regime of photon
 virtualities in the 
inclusive deep-inelastic electron-proton scattering process, a definition
for the total cross section for virtual-photon/proton interactions is required.
It is obtained by adopting a convention for the virtual-photon flux
from the electron beam~\cite{pr_167_1365,*pr_120_1834} in order to relate
the $\gamma^* p$ cross section to the $ep$ cross section. Such a definition
makes physical sense if the lifetime of the virtual photon is sufficiently
 long that the photon traverses two proton radii. One can show
that this condition corresponds to values for the Bjorken scaling
 variable {\xbj} less than 0.06.\hspace*{-0.7ex}~\cite{hardproc,pr_54_3194,stmp_140} In this low-$x$
region, $x$ may be approximated by $\xbj\simeq\qsq /W^2$\hspace*{-1.5ex},\hspace*{1.ex} and the total cross section is related to
the proton structure function ${\fsq}({\xbj},\qsq)$ via
\begin{eqnarray}
\label{eq:virtot}
\sigma_{{\gamma^*p}}^T + \sigma_{{\gamma^*p}}^L &\approx& \frac{4\pi^2\alpha}{\qsq (1-\xbj)} {\fsq}({\xbj},\qsq),
\end{eqnarray}
where the photon-proton cross section has been expressed as the sum
of its contributions from longitudinal and transverse photons, and $\alpha$ is
the fine structure constant. Figure~\ref{fig:sigtot}
compares the photon-proton total cross sections for real and virtual photons
over the entire region covered by fixed-target experiments and by the collider
experiments at HERA, along with a dashed line representing the condition 
$\xbj = 0.06$. At high photon virtuality, the HERA results extend from 
the region
where this condition no longer holds, and the cross section falls rapidly with
increasing {\xbj}, to the region where the cross section is steeply rising
with energy, driven by the gluon density in the proton. The total cross
section for real photons at high energy exhibits the weak energy dependence
characteristic of models based on Regge theory. The dearth of measurements in
the kinematic region to be covered by the EPIC collider clearly shows the need
for further measurements in this region of transition between nonperturbative
and perturbative behavior.
\section{Diffractive contribution to the total {$\gamma^* {\rm p}$} cross section}
\begin{figure}[htbp]
\begin{center}
\epsfig{file=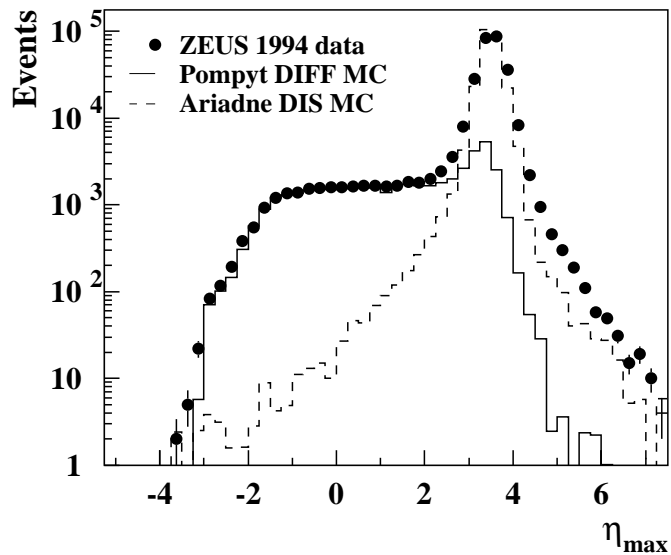, width=0.75\textwidth, bbllx=11, bblly=6,
  bburx=233, bbury=189, clip=}
\caption{ 
\label{fig:etamax_pompyt}
Sample distribution in $\eta_{\rm max}$, the rapidity of the most forward
energy deposit, compared to simulations distinguishing the diffractive
component in the inclusive electroproduction process
} 
\end{center}
\end{figure}

The initial observation of a hard diffractive process at HERA was made during
the early analyses of the proton structure function {\fsq}. Data
selection criteria related to the forward direction were found to reject
a category of events which were not present in the available 
simulations. An event candidate exemplifying this 
subcategory of the total photon-proton cross section is shown in
\begin{figure}[htbp]
\begin{center}
\epsfig{file=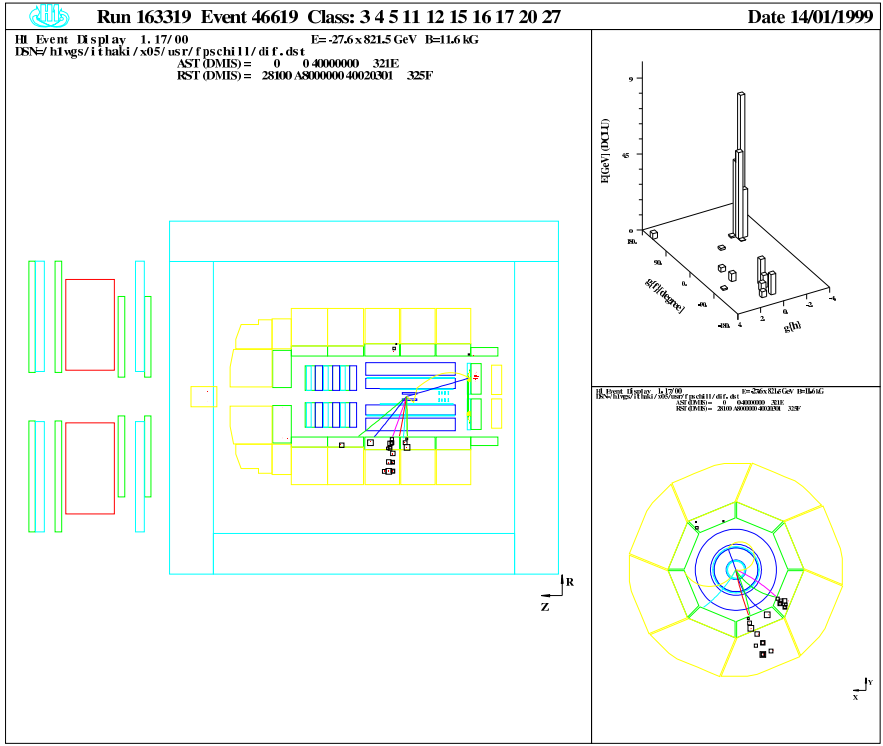, width=0.83\textwidth, angle=90}
\caption{ 
\label{fig:h1event}
Event recorded by the H1 detector 
exhibiting a diffractive topology with a rapidity gap
between the photon fragmentation region and the forward direction
} 
\end{center}
\end{figure}
Fig.~\ref{fig:h1event}.
These events exhibit a rapidity gap between the most forward energy deposit
in the central detector and the proton flight direction. 
A distribution
in this event variable, dubbed $\eta_{\rm max}$, is shown in
Fig.~\ref{fig:etamax_pompyt}, together with the results of simulations
of the diffractive (POMPYT) and nondiffractive (ARIADNE) components.
The absence of event 
activity between the target and photon fragmentation regions
is used to characterize event topologies arising from this inclusive
diffractive
process, since it indicates the absence of color flow and hence the exchange
of a colorless object as the underlying production mechanism.

This means of isolating a clean sample of diffractive events is supplemented
in the ZEUS~\cite{ichep98_789} and H1~\cite{epj_6_587} experiments by trigger
selection criteria based on forward baryon detectors. The identification
of a forward proton or neutron provides a very clean technique for isolating
the diffractive component. Measurement of the baryon momentum provides further
information on the dissociation of the initial-state proton. Each of the
experiments operates both a forward proton spectrometer and a forward neutron
calorimeter. Such detector components will be essential for the study of
diffractive physics at the EPIC collider, since the rapidity-gap selection 
technique
will suffer from the limited rapidity range at the lower energy. 
The strict requirements on such detectors, which include extreme radiation
conditions, very accurate momentum measurement, and excellent angular
resolution at very small production angles impose the necessity of close
interaction with the accelerator physicists during both the design and
operation phases. For example, the calibration of the ZEUS leading-proton
spectrometer, which consists of silicon-strip detectors in Roman pots, 
depends crucially on accurate knowledge of the geometry of
beam-line components and their magnetic fields, and its operation is possible
only under clean background conditions which are very sensitive to
the proton-beam injection procedure. The design of experiments for diffractive
physics at the EPIC collider will thus profit greatly 
from the experience gained at HERA.
\section{Exclusive processes}
The data obtained by the HERA collider experiments during their first
six years of operation have proven them to be excellent sources of information
on exclusive strong processes.
Such investigations  cannot be performed
at $pp$ or {\ppbar} colliders, where the acceptance of the 
central detector components is limited to interactions 
producing high transverse energy in the final state, 
or at $e^+e^-$ colliders, where the
primary interaction is electroweak. The electron-proton collider geometry
has thus been shown to result in sensitivity to a rich variety of
strong-interaction phenomena inaccessible by other means.
The excellent solid-angle coverage of the H1 and ZEUS central 
detectors, along with simple special-purpose scintillation counters
in the forward region, permits selection of the exclusive reaction
$ep \rightarrow epV$, where $V$ is a vector meson ({\rhoz}, $\omega$,
$\phi$, {\jpsi}, $\Upsilon$), with the background from proton-dissociative
processes limited to less than 20\%~\cite{stmp_140}\hspace*{-6pt}.\hspace*{6pt} (The latter semi-exclusive
process has sparked much theoretical interest recently and will be discussed in
Sect.~\ref{sec:semi-exclusive}.) Of particular interest to
instrumentation experts is the necessity of a low-noise nearly-hermetic
central calorimeter to provide high-resolution event criteria for exclusivity
in such studies. These diffractive processes offer three variables as 
tunable gauges for the scale of the interaction: {\qsq}\hspace*{-1.2ex},\hspace*{1.2ex} $M_V$ and $t$,
and thus hold great promise for tests of QCD models of diffraction.
The kinematically required minimum transverse kick means that the
parton density functions are necessarily skewed, with the 
skewedness parameter,
$\delta\equiv (x_1-x_2)$, given by
\begin{eqnarray}
|t|_{\rm min} &=& m_{\rm p}^2 \, \frac{(M_V^2 + \qsq)^2}{W^4} \;=\; m_{\rm p}^2 \, \delta^2.
\end{eqnarray}
It has  been shown recently that the comparison of exclusive
{\jpsi} and {$\Upsilon$} photoproduction cross sections yields information
on the evolution of the skewed gluon density in the
proton.\hspace*{-0.7ex}~\cite{pl_454_339,hep98_12_316} In the following, we give examples of
HERA measurements which demonstrate the choice of scale parameter by
distinguishing data samples of low and high photon virtuality.
\subsection{Exclusive photoproduction of vector mesons}
\label{sec:exphotoproduction}
\begin{figure}[ht]
\vspace*{-4mm}
\begin{center}
\epsfig{file=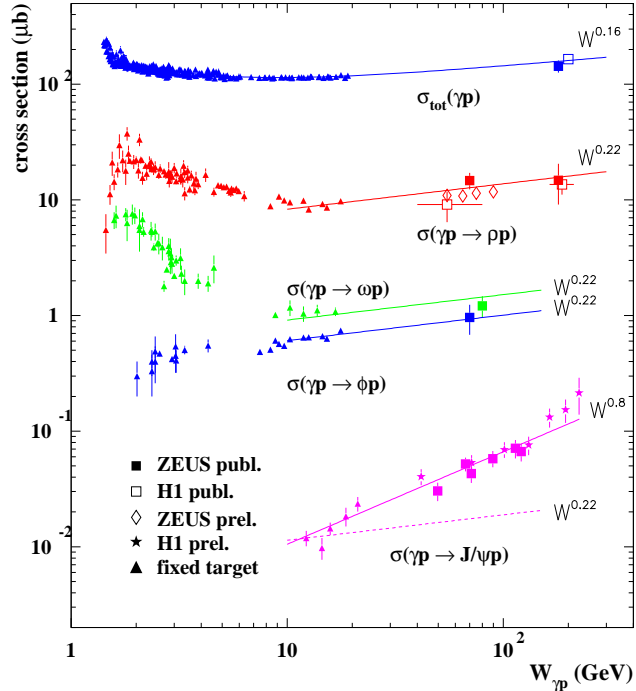, width=0.7\textwidth, bbllx=0, bblly=0, bburx=570, bbury=626, clip=}
\vspace*{-4mm}
\caption{ 
\label{fig:phpvm}
Measurements of the exclusive photoproduction cross sections for {\rhoz},
$\omega$, $\phi$ and {\jpsi} mesons versus energy, compared to the
total $\gamma p$ cross section. The lines indicate the slopes of three types of power-law dependence
} 
\end{center}
\end{figure}

The central detectors of the H1 and ZEUS experiments are efficient for the
detection of the scattered electron for photon virtualities exceeding about
$4~{\gev}.\hspace*{-0.7ex}^2$ A simple signature for the exclusive photoproduction of vector
mesons is thus the reconstruction of the decay products in the central
detector in the absence of any other energy deposits. 
Indeed, the first exclusive reaction cross section measured at HERA was that
for
the
photoproduction of {\rhoz} mesons~\cite{zfp_69_39}\hspace*{-10pt},\hspace*{10pt}
owing to its large cross section
(10~{$\mu$}b, about 7\% of the total photon-proton cross section)
and simple event topology. These measurements, as well as those that followed
for the $\phi$ and $\omega$ mesons, exhibited the weak energy dependence
and steeply forward peaking characteristic to soft diffractive processes.
Figure~\ref{fig:phpvm} 
shows the energy dependence of the exclusive photoproduction cross
sections for the light vector mesons, comparing it to that of the total photon-proton
cross section and to
that observed in exclusive {\jpsi} photoproduction.\hspace*{-0.7ex}~\cite{eps99_157aj}
The power of 0.32 (see Sect.~\ref{sec:phenomenology}), expected for
the forward cross sections according to Regge phenomenology, is reduced
to  0.22 by the integration over $t$ and 
the energy dependence in the Pomeron trajectory.
The much steeper energy dependence observed in {\jpsi} photoproduction 
is consistent with the prediction of Ryskin~\cite{zfp_57_89} and has
inspired efforts to extract the gluon density in the proton from these
measurements~\cite{zfp_76_231,*pr_57_512}\hspace*{-9pt}.\hspace*{9pt} The measurements from fixed-target
experiments at $W\simeq 10~{\gev}$ indeed served as the initial indication
that the energy dependence was strong but the kinematics and statistics of
those measurements precluded  a variety of further studies ($t$ dependence,
decay-angle analyses) which will be made possible by the EPIC collider.

\begin{figure}[htbp]
\begin{center}
\epsfig{file=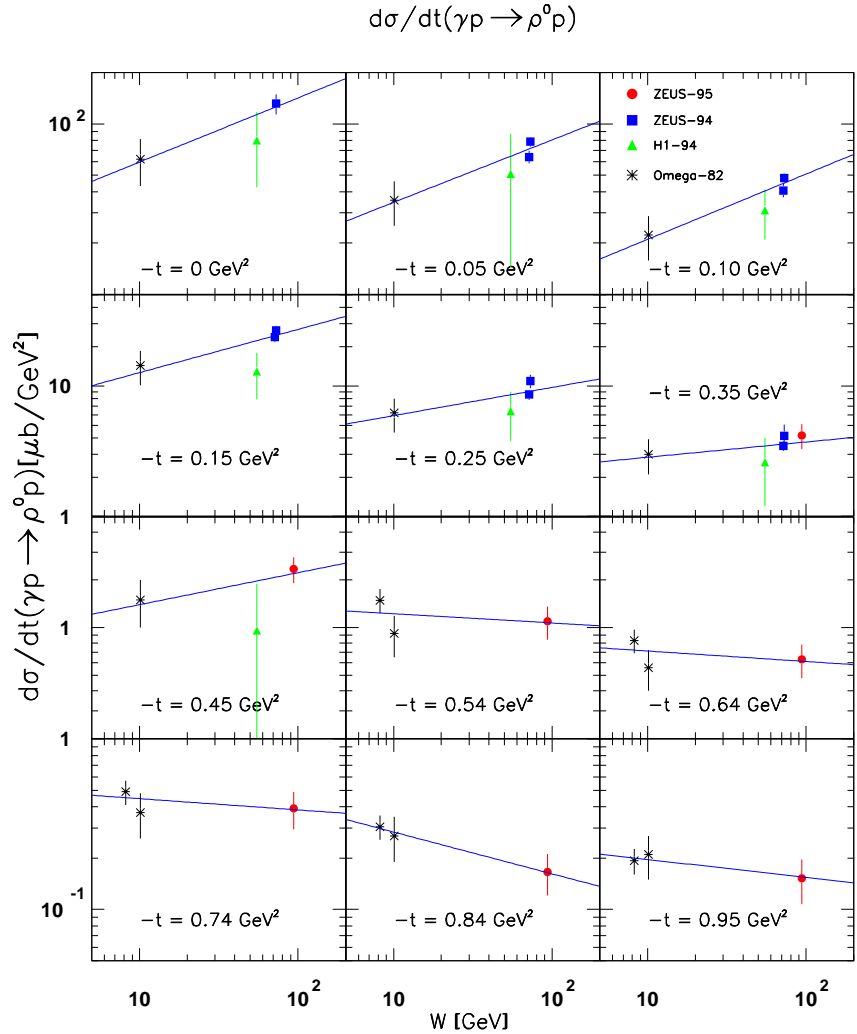, width=0.95\textwidth, bbllx=19, bblly=112, bburx=528,  bbury=727, clip=}
\vspace*{-4mm}
\caption{
\label{fig:new-dsdtrho} 
Exclusive {\rhoz} photoproduction cross sections as a function of the
photon-proton center-of-mass energy, $W$, for
various values of $t$. The analysis compares the measurements by the H1~\protect\cite{np_463_3} and ZEUS~\protect\cite{ichep98_788,epj_2_247,zfp_73_253} collaborations
at high energy to those performed at low energy by the OMEGA
experiment~\protect\cite{np_209_56}\hspace*{-9pt}.\hspace*{9pt} The error bars show the quadratic sum of
statistical and systematic uncertainties.
The lines indicate the results of fits to the form $\frac{\D\sigma}{\D
  t}\propto (W^2)^{2\alpha (t) -2}$ performed in order to determine the
parameters of the Pomeron trajectory \vspace*{-4mm}
}
\end{center}
\end{figure}

\begin{figure}[ht]
\begin{minipage}[t]{0.48\textwidth}
\begin{center}
\epsfig{file=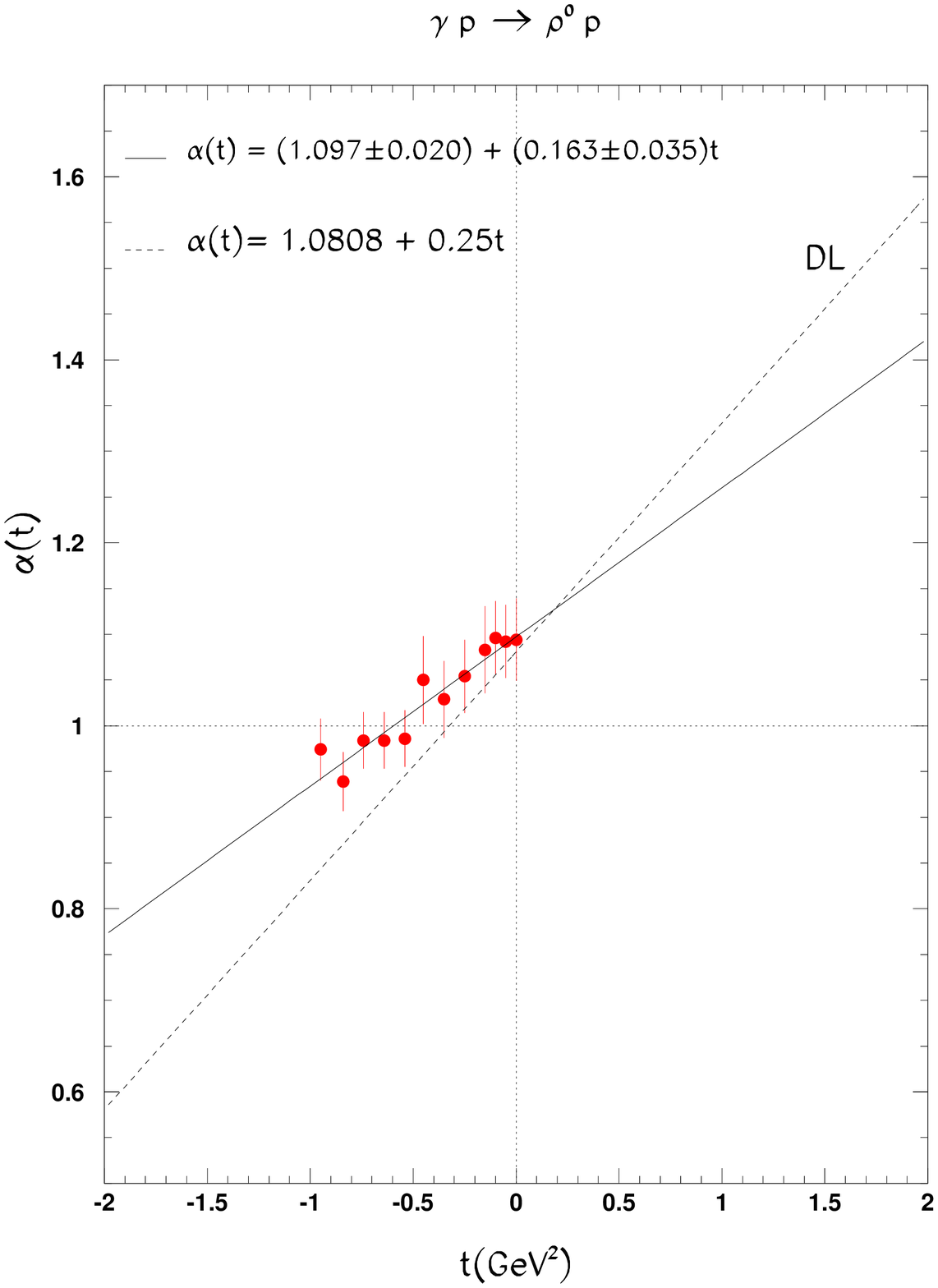, width=0.9\textwidth, bbllx=16, bblly=67, bburx=524, bbury=764, clip=}
\end{center}
\end{minipage}
`\hfill
\begin{minipage}[t]{0.48\textwidth}
\begin{center}
\epsfig{file=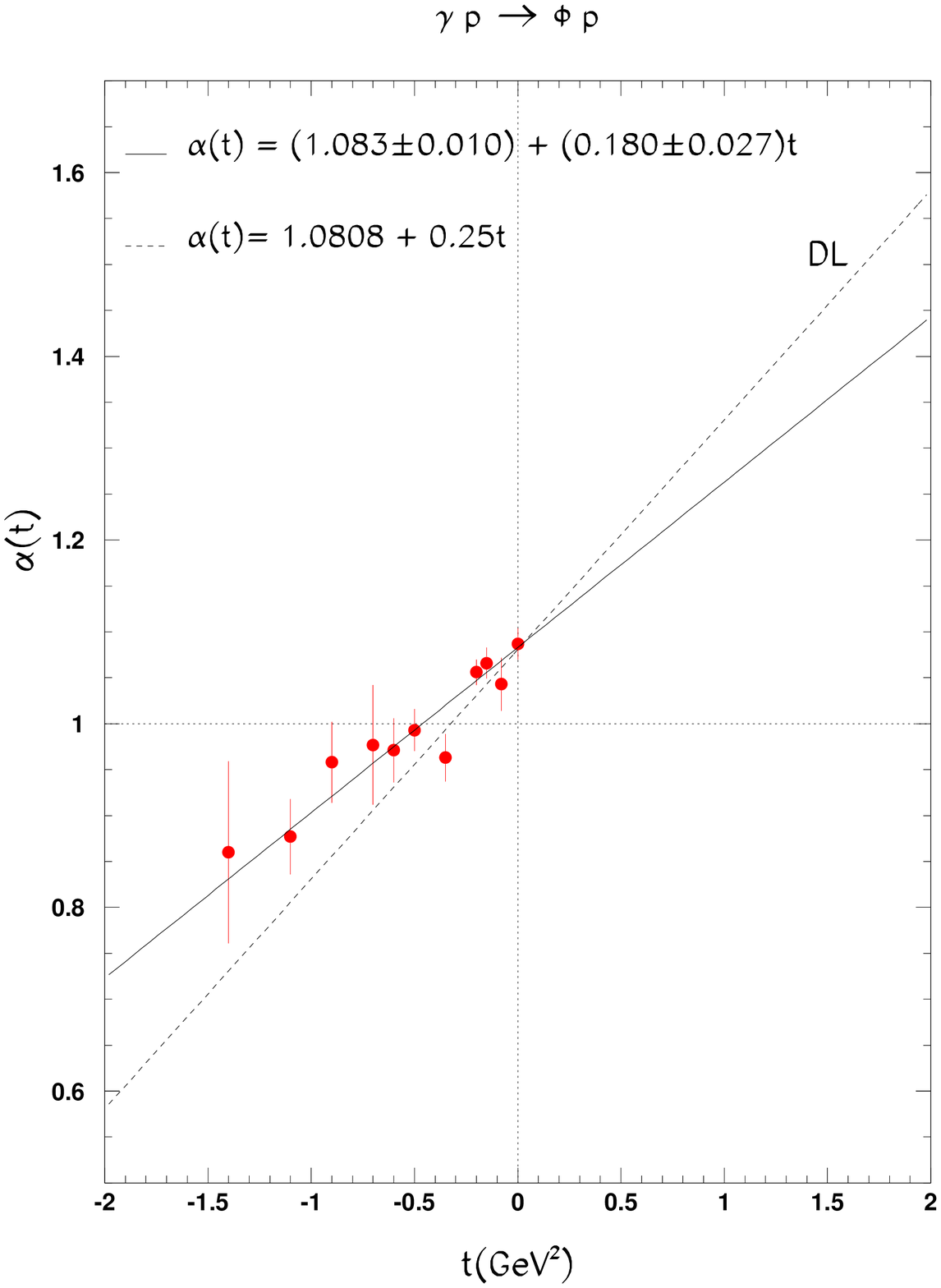, width=0.9\textwidth, bbllx=16, bblly=67, bburx=524, bbury=764, clip=}
\end{center}
\end{minipage}
\vspace*{-4mm}
\caption{ 
\label{fig:new-a}
Preliminary results for the determination of the Pomeron trajectory in the
exclusive photoproduction of 
{\rhoz} and $\phi$ mesons by the ZEUS collaboration~\protect\cite{ichep98_788}\hspace*{-9pt}.\hspace*{9pt} The error bars show the statistical uncertainties
in the fits to the HERA data and the low-energy data as shown for the {\rhoz}
in Fig.~\protect\ref{fig:new-dsdtrho}. The result
of a fit to a linear trajectory (solid line)  is compared to the 
trajectory determined by Donnachie and Landshoff using
hadronic interaction cross sections~\protect\cite{pl_296_227} (dashed line)
} 
\end{figure}

As mentioned in Sect.~\ref{sec:phenomenology}, one means of characterizing
the underlying production process in the exclusive photoproduction of vector
mesons is given by studying the energy dependence of the forward peaking.
In the framework of Regge phenomenology, such a dependence arises from
the $t$ dependence of the Pomeron trajectory: $\alpha (t) = 1.08 + 0.25~t$.
Such a dependence is not expected in models which assume the exchange of
a pair of gluons in a color-singlet state.\hspace*{-0.7ex}~\cite{shep_11_51} Its study
requires high experimental sensitivity, since the dependence on energy 
is logarithmic. At present, the range of energy available to the H1 and ZEUS
experiments has not sufficed to accurately measure such shrinkage effects,
motivating attempts to include low-energy data in the
analysis~\cite{pl_424_191,ichep98_788}\hspace*{-19pt}.\hspace*{19pt} Figure~\ref{fig:new-dsdtrho}
shows the result of such an investigation by the ZEUS
collaboration~\cite{ichep98_788} for exclusive photoproduction of {\rhoz}
mesons. 
By restricting the low-energy data to that of the OMEGA
experiment~\cite{np_209_56} in order to exclude data at $W<8~\gev$, which
suffer from contamination of the Pomeron exchange mechanism by other exchange
mechanisms~\cite{hep99_08_218}\hspace*{-2.4ex},\hspace*{2.4ex} this analysis achieves a clear measurement of the $t$ dependence,
covering a $t$ range in which the energy dependence evolves from rising to
falling. The Pomeron trajectories determined for the {\rhoz} and {$\phi$}
via this method are shown in Fig.~\ref{fig:new-a}, along with the results of
fits illustrating the type of $t$ dependence observed. 
The observed $t$ dependence is weaker than that extracted from
the Donnachie-Landshoff parametrization of hadronic cross
sections~\cite{pl_296_227,pl_231_189}\hspace*{-19pt}.\hspace*{19pt} Of particular interest will be
the extension of these investigations to higher momentum transfer,
in order to identify a transition to a regime where perturbative exchange
mechanisms apply. For such studies, further measurements at $W\gsim 10~\gev$
are necessary. Furthermore, an initial analysis of this effect
in {\jpsi} photoproduction has yielded the conclusion that
such a $t$ dependence is absent and thus that the underlying exchange
mechanism is consistent with the perturbative expectation and inconsistent
with Regge phenomenology~\cite{pl_424_191}\hspace*{-10pt}.\hspace*{10pt} Unfortunately, the lack
of low-energy data required the use of measurements performed using
a muon beam on an iron target~\cite{np_213_1}\hspace*{-10pt},\hspace*{10pt} rendering the result
controversial, owing to uncertainties associated with
nuclear effects~\cite{ichep98_572}\hspace*{-10pt}.\hspace*{10pt}  Investigations restricted to the HERA energy regime
alone lack the sensitivity to rule out the shrinkage effect at the 
level observed for the light vector mesons~\cite{eps99_157aj}\hspace*{-10pt}.\hspace*{10pt}
Here again, a clear understanding of this issue, and 
the associated measurement program to further investigate the dynamics
of the production process require measurements in the EPIC energy range.
\subsection{Exclusive electroproduction of vector mesons}
\begin{figure}[ht]
\begin{center}
\epsfig{file=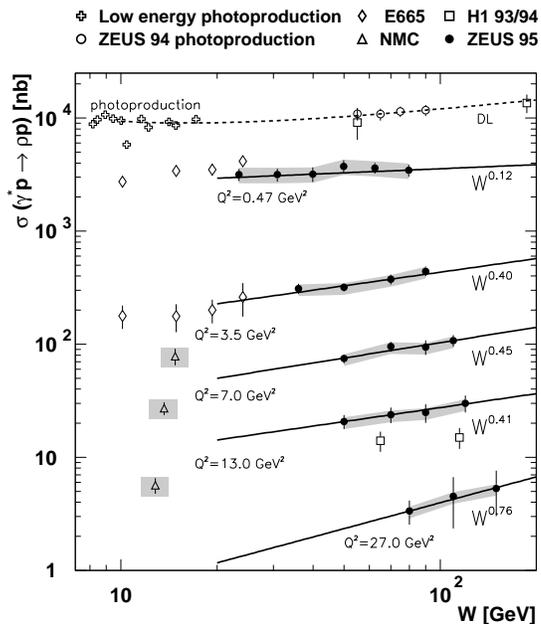, width=0.6\textwidth, bbllx=14, bblly=27,
  bburx=428, bbury=504, clip=}
\vspace*{-4mm}
\caption{   
\label{fig:allvm}
Comparison of cross sections for  exclusive $\rho^0$ electroproduction
as a function of $W$ for various values of {\qsq}\hspace*{-1.2ex}.\hspace*{1.2ex} 
The error bars represent statistical and systematic errors added in quadrature.
The solid lines represent results of fits to the ZEUS data~\protect\cite{epj_6_603}\hspace*{-2.5ex}.\hspace*{2.5ex}
The dashed line is the prediction for the total photon-proton cross section 
by Donnachie 
and Landshoff~\protect\cite{pl_296_227}\hspace*{-2.5ex}.\hspace*{2.5ex}
The overall normalization uncertainties are shown as shaded bands for the
NMC and ZEUS data points.
The NMC~\protect\cite{np_429_503}\hspace*{-2.5ex},\hspace*{2.5ex}  E665~\protect\cite{zfp_74_237}
and H1~\protect\cite{np_468_3}  data points  
were interpolated  to the indicated  $Q^2$ values  
} 
\end{center}
\end{figure}

Investigation of the exclusive electroproduction of {\rhoz} mesons 
provides information on the r\^ole of the photon virtuality in establishing
a hard scale, since we know from measurements described in the 
previous section that {\rhoz} photoproduction at low {\abst} can be
described by a soft diffractive production mechanism. For example, 
the proposed perturbative description of this process implies an energy
dependence which increases with photon virtuality, following the behavior
of the gluon density in the proton as measured in the inclusive $ep$
scattering
process.
Accurate measurements at high {\qsq} have been difficult at HERA 
owing to statistical limitations. The H1~\cite{np_468_3,dr_99_010} and
ZEUS~\cite{epj_6_603} collaborations have 
published measurements based on samples of a few thousand 
events. Figure~\ref{fig:allvm}
shows the energy dependence of the cross sections for exclusive {\rhoz}
electroproduction at various values of {\qsq}\hspace*{-1.2ex},\hspace*{1.2ex} comparing the fixed-target
measurements at low energy~\cite{np_429_503,zfp_74_237} to those by the
H1~\cite{np_468_3}\footnote{The H1 data shown here have since been superseded by
their more recent measurements.\hspace*{-0.7ex}~\protect\cite{dr_99_010} 
The new results are
without consequence for our present discussion of the need for more
measurements at low energy.} 
and ZEUS~\cite{epj_6_603} collaborations. 
The
important question of whether the energy dependence steepens with increasing
{\qsq} cannot be definitively resolved at the present level of statistical
accuracy. Worse yet, this QCD prediction actually 
holds only for the longitudinal
cross section,\hspace*{-0.7ex}~\cite{pr_50_3134,pr_56_2982} whereas here we consider the sum of 
longitudinal and transverse cross sections. Extraction of the longitudinal
contribution requires the measurement of 
$R\equiv{\sigma_{\rm L}}/{\sigma_{\rm T}}$ 
(to be described in Sect.~\ref{sec:helicity}),
introducing an additional source of uncertainty.\hspace*{-0.7ex}~\cite{epj_6_603} The 
desired increase in 
sensitivity to the energy dependence obtained by the inclusion
of the low-energy measurements suffers from an apparent discrepancy between
the results at ${\qsq}=3.5\,\gevsq$ from the E665 and NMC collaborations.
More accurate measurements at low energy, of sufficient precision to
accurately extract the longitudinal cross section, are of paramount importance
for progress on this issue.

The exclusive electroproduction of {\jpsi} mesons has been investigated
recently by both the H1~\cite{dr_99_026} and ZEUS~\cite{epj_6_603} 
collaborations in order to shed light on the relationship between {\qsq} and
$M_{J/\psi}$ in determining the scale of the reaction. While experiments
at the EPIC collider will not attain the high {\qsq} values reached by the HERA
experiments, the high luminosity at EPIC will permit accurate measurements
of {\jpsi} electroproduction
well into the perturbative regime, thus providing together with the results
from HERA the long lever arm in energy which is essential for the
understanding of the diffractive production process.

\subsection{Helicity analyses}
\label{sec:helicity}
The simplicity of the final state in these exclusive processes permits
detailed investigations of the decay-angle distributions, which have provided
much
 information on the helicity structure of the diffractive production
mechanism. The limited extent of the accessible range in $y$ precludes the separation of
longitudinal and transverse cross sections unless the proton beam energy
is varied,\footnote{This will also be the case at the EPIC collider.} which
has not yet been done at HERA. Thus the analysis of the decay-angle
distributions has been employed as a means of measuring $R$. This type of 
analysis provided another early success of the perturbative QCD models
of exclusive diffractive processes by verifying the prediction that the longitudinal
cross section exceed the transverse cross section at high {\qsq}\hspace*{-1.2ex}.\hspace*{1.2ex} 

Three angles suffice to completely describe the exclusive electroproduction
of vector mesons, as shown in Fig.~\ref{fig:defangles}: the azimuthal angle between the scattering plane and the production plane, 
$\Phi_h$, and the two $\rho^0$ decay angles: 
\begin{figure}[htbp]
\begin{center}
\epsfig{file=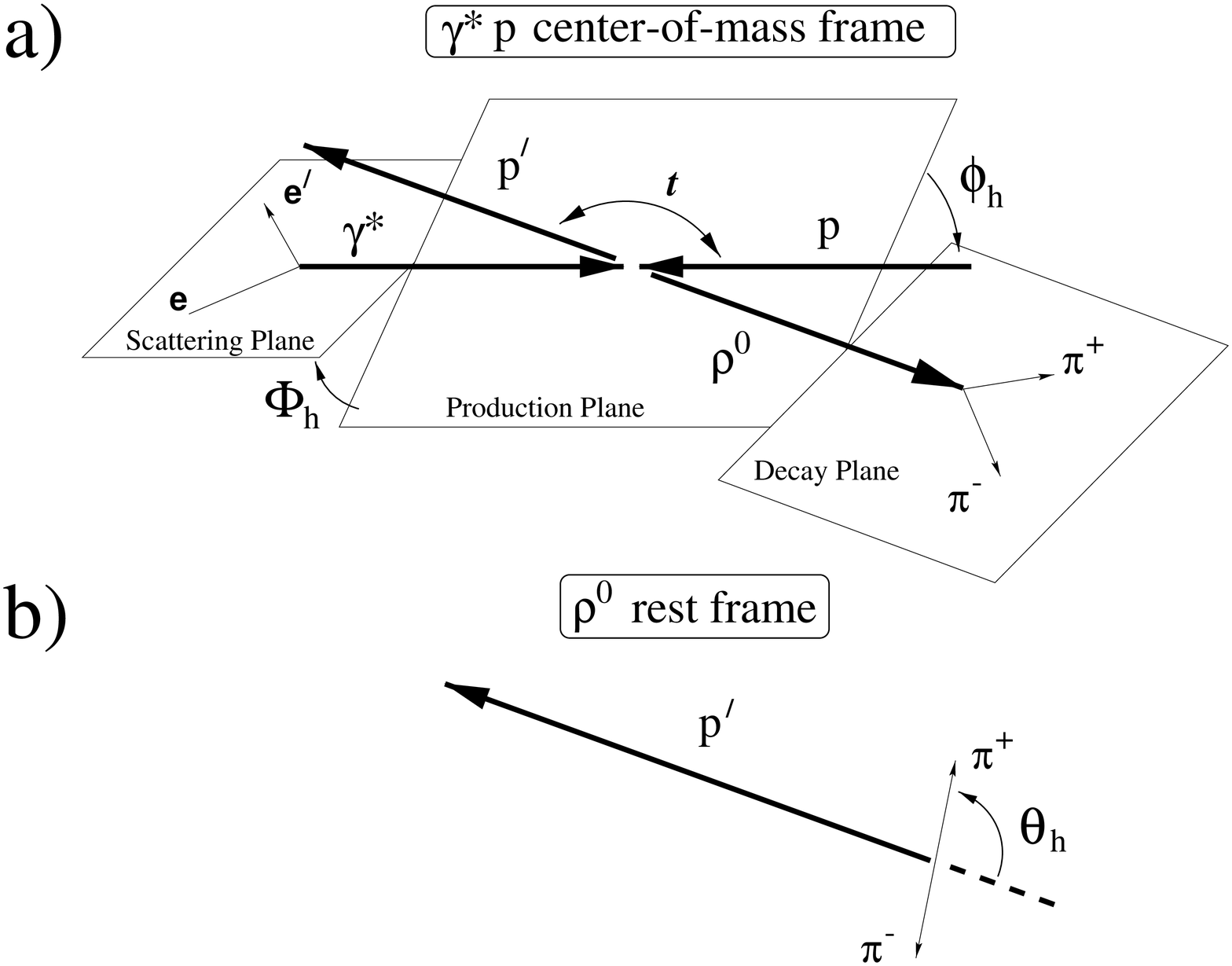, width=0.68\textwidth}
\vspace*{-4mm}
\caption{
\label{fig:defangles} 
Schematic diagram of the process $ep \rightarrow e \rho^0 p$ (a) in the
virtual-photon/proton center-of-mass system, and (b) in the rest frame of
the $\rho^0$ meson\vspace*{-2mm}
}
\end{center}
\end{figure}
$\phi_h$, the azimuthal angle 
between the production and decay planes, defined in either 
the virtual-photon/proton system or in
the $\rho^0$ rest frame, and $\theta_h$, which is the polar angle of the 
positively charged decay product, defined with respect to the direction
of the $\rho^0$ momentum vector in the virtual-photon/proton system, which is 
the same direction as that opposite to the momentum vector of the final-state
proton in the rest frame of the $\rho^0$ meson. This latter choice of 
spin-quantization axis defines the helicity frame, in which helicity
conservation was found to hold approximately in exclusive {\rhoz} photoproduction
experiments at SLAC in the early 1970s.\hspace*{-0.7ex}~\cite{pr_5_545,pr_7_3150}  

\begin{figure}[htbp]
\begin{minipage}[t]{0.48\textwidth}
\begin{center}
\epsfig{file=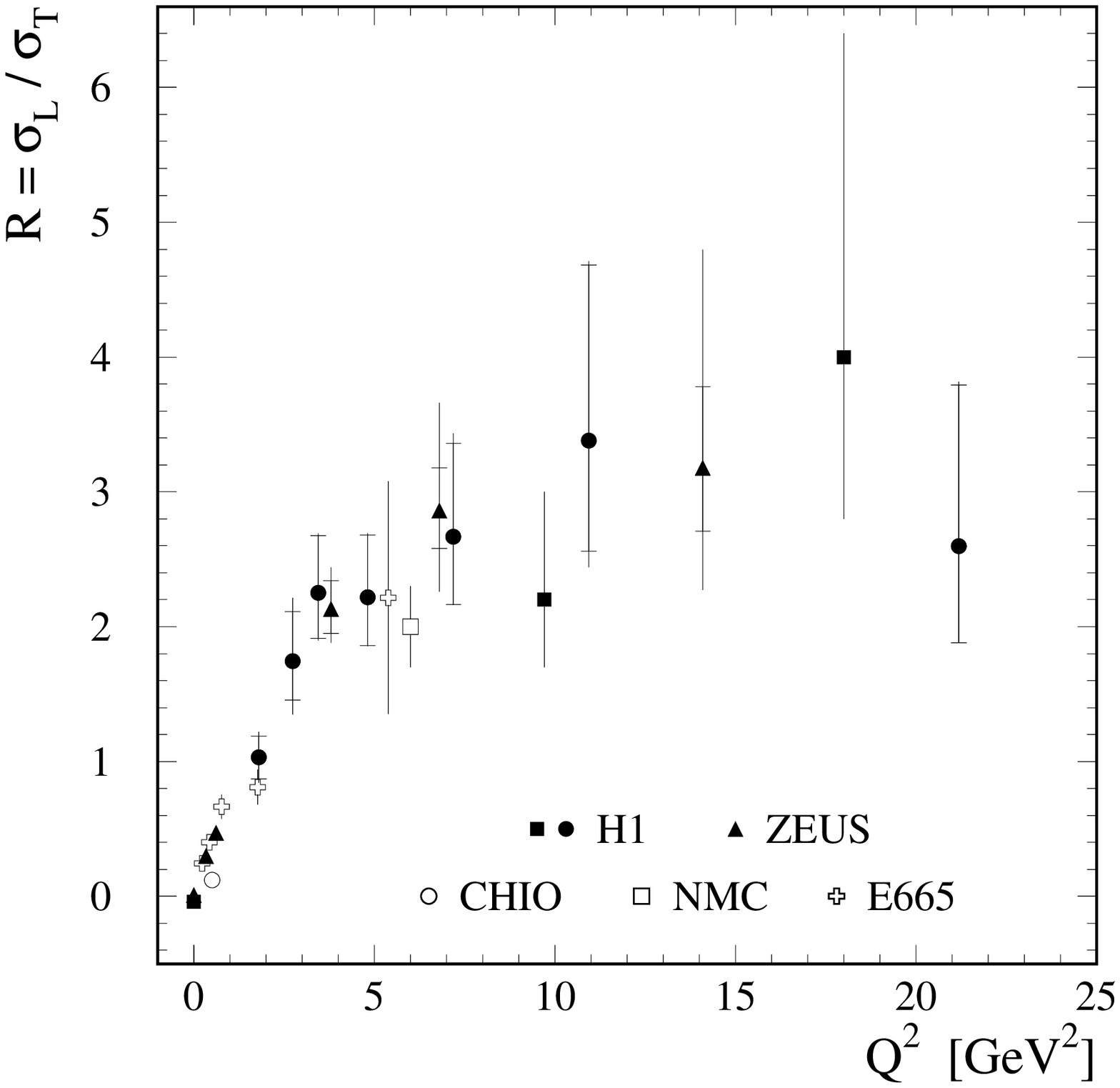, width=\textwidth}
\caption{ 
\label{fig:h1_R}
Measurements of the ratio of exclusive {\rhoz} electroproduction cross sections
for longitudinal and transverse photons, $R$, as a function of photon
virtuality, {\qsq}
} 
\end{center}
\end{minipage}
\hfill
\begin{minipage}[t]{0.48\textwidth}
\begin{center}
\epsfig{file=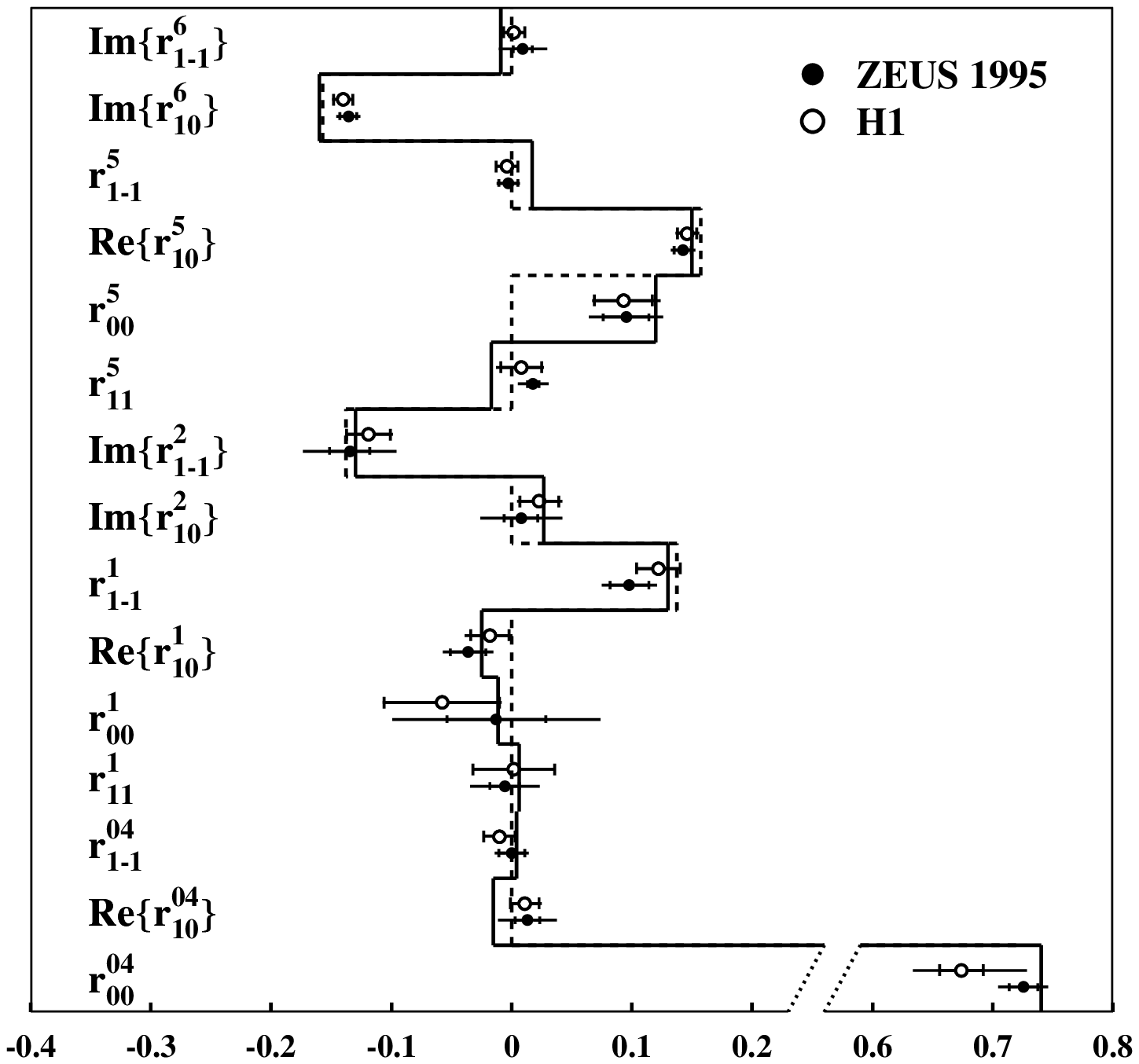, width=\textwidth, bbllx=43, bblly=37,
  bburx=462, bbury=426, clip=}
\caption{ 
\label{fig:me_dis}
Combinations of spin-density matrix elements measured for {\rhoz}
electroproduction in the kinematic region \mbox{$3<{\qsq}<30~\gevsq$},
\mbox{$40 < W <120~\gev$} and \mbox{${\abst}<0.6~\gevsq$} by the ZEUS
collaboration.\hspace*{-0.7ex}~\protect\cite{dr_99_102} See text for full description
} 
\end{center}
\end{minipage}
\end{figure}

Following the work of Schilling and Wolf,\hspace*{-0.7ex}~\cite{np_61_381} 
the three-dimensional angular distribution has been parametrized as follows:
\begin{eqnarray}
W({\theta_h},\phi_h,\Phi_h) &\hspace*{-1pt}=&\hspace*{-1pt} \frac{3}{4\pi} \biggl[ 
\frac{1}{2}(1-r^{04}_{00})+\frac{1}{2}(3r^{04}_{00}-1)\cos^2{\theta_h} \nonumber\\
& &\hspace*{-25pt}-\sqrt{2}~{\rm Re}\{r^{04}_{10}\}\sin{2\theta_h}\cos{\phi_h} -r^{04}_{1-1}\sin^2{\theta_h}\cos{2\phi_h} \nonumber\\
& &\hspace*{-25pt}-\epsilon\cos{2\Phi_h}(r^1_{11}\sin^2{\theta_h}+r^1_{00}\cos^2{\theta_h}-\sqrt{2}~{\rm Re}\{r^1_{10}\}\sin{2\theta_h}\cos{\phi_h}\nonumber\\
& &\hspace*{-25pt}\hspace*{1.8cm}-r^1_{1-1}\sin^2{\theta_h}\cos{2\phi_h}) \nonumber\\
& &\hspace*{-25pt}-\epsilon\sin{2\Phi_h}(\sqrt{2}~{\rm Im}\{r^2_{10}\}\sin{2\theta_h}\sin{\phi_h}+{\rm Im}\{r^2_{1-1}\}\sin^2{\theta_h}\sin{2\phi_h})\nonumber\\
& &\hspace*{-25pt} +\sqrt{2\epsilon (1+\epsilon)}\cos{\Phi_h} (r^5_{11}\sin^2{\theta_h}+r^5_{00}\cos^2{\theta_h} \nonumber\\
& &\hspace*{-25pt}\hspace*{2.cm}-\sqrt{2}~{\rm Re}\{r^5_{10}\}\sin{2\theta_h}\cos{\phi_h} -r^5_{1-1}\sin^2{\theta_h}\cos{2\phi_h})\nonumber\\
& &\hspace*{-25pt} +\sqrt{2\epsilon (1+\epsilon)}\sin{\Phi_h} (\sqrt{2}~{\rm Im}\{r^6_{10}\}\sin{2\theta_h}\sin{\phi_h} \nonumber\\
& &\hspace*{-25pt}\hspace*{3.7cm}+{\rm Im}\{r^6_{1-1}\}\sin^2{\theta_h}\sin{2\phi_h}) \biggr],
\label{full_equation}
\end{eqnarray}
where the superscripts of the combinations of spin-density matrix elements
correspond to the helicity degrees of freedom of the virtual photon, and
the subscripts to those of the dipion state. 
When the dipion state is spin 1, the fifteen coefficients   $r^{04}_{ik}$,
$r^{\alpha}_{ik}$ are 
related directly to various combinations of
the helicity amplitudes, $T_{\lambda_{\rho} \lambda_{\gamma}}$, where 
$\lambda_{\rho}$ and $\lambda_{\gamma}$ are the helicities of the $\rho^0$
meson and of the photon, respectively. The assumption that helicity is
conserved in the photon/vector-meson transition when the amplitudes
are defined in the helicity frame ({"s-channel helicity conservation"} or 
"SCHC"),
with the consequence that the degree of {\rhoz} polarization is equal
to the ratio of the longitudinal and transverse cross sections, allows the
extraction of this ratio $R$ from the distribution in polar angle alone. A compilation
of determinations of $R$ via this method is shown in Fig.~\ref{fig:h1_R}. 
The data from the fixed-target muon experiment at FNAL, E665,\hspace*{-0.7ex}~\cite{zfp_74_237}
and the low-{\qsq} data from the ZEUS collaboration identify a region
of transition to increasing values of $R$, with the longitudinal cross section becoming
dominant for $\qsq\gsim 2~\gevsq$\hspace*{-1.2ex}.\hspace*{1.2ex} The latest results from the H1
collaboration~\protect\cite{dr_99_010} indicate that the steep rise observed
at intermediate values of {\qsq} does not persist at high {\qsq}\hspace*{-1.2ex}.\hspace*{1.2ex}

The ZEUS~\cite{dr_99_102} and H1 collaborations~\cite{dr_99_010} 
have recently completed analyses
of the three-dimensional angular distribution for data samples of a few
thousand events, extracting the fifteen
coefficients $r^{04}_{ik}$, $r^{\alpha}_{ik}$. Figure~\ref{fig:me_dis}
shows the results from the ZEUS collaboration 
in the kinematic region \mbox{$3<{\qsq}<30~\gevsq$}\hspace*{-1.2ex},
\mbox{$40 < W <120~\gev$} and \mbox{${\abst}<0.6~\gevsq$}\hspace*{-1.2ex}, 
comparing them to the results
from the H1
collaboration in a similar kinematic region, 
and to a calculation by
Ivanov and Kirschner~\protect\cite{pr_58_114026} (solid line). 
The inner error bars represent the statistical
uncertainty; the outer error bars show the quadratic sum of
statistical and systematic uncertainties. 
Also shown
are the values of the coefficients predicted according to the hypothesis 
of helicity
conservation in the $s$-channel amplitudes~\cite{np_61_381} (dashed line).
Of particular interest is the violation of SCHC evinced in the nonzero value
for the coefficient
$r^5_{00}$, which is manifested 
in the distribution in the azimuthal angle between the
scattering and production planes ({\Phih}). This observation implies
a small nonzero single-flip amplitude for the production of {\rhoz} mesons in 
helicity state 0 from transverse photons. Such a violation has now been
reproduced in two further recent calculations based on differing
assumptions.\hspace*{-0.7ex}~\cite{schnc_dis99} These measurements show the level of
violation of SCHC to be small enough that the effect of the assumption of
SCHC in the earlier determinations of $R$ is much smaller than the other
sources of uncertainty in those determinations.
The ZEUS collaboration also performed the 
analysis in the low-{\qsq} region,\hspace*{-0.7ex}~\cite{dr_99_102} finding
a value for $r^5_{00}$ of similar magnitude, but also distinct indications
for a more complex pattern of helicity violation, including double-flip
contributions, as found in {\rhoz} photoproduction (see Sect.~\ref{sec:semihelicity}).

In addition to the higher luminosity, which will allow such detailed
decay-angle analyses also for {\jpsi} and {$\Upsilon$} mesons, experiments at
the EPIC collider will have two distinct advantages over the 
investigations at HERA. First, the longitudinal
polarization of the electron beam will permit measurement of spin-density
matrix elements currently inaccessible at HERA (namely, those of
superscript~3, corresponding to circular polarization of the photon).
The E665 collaboration, profiting from the longitudinal polarization of the
muon beam, have published a measurement~\cite{zfp_74_237} of such coefficients at low
{\qsq}\hspace*{-1.2ex}.\hspace*{1.2ex} The HERA collider experiments may provide such
measurements once the necessary spin rotators have been installed and the 
polarization program established, as is now scheduled to occur in 2001/2.
The second major advantage of research at the EPIC collider is that the the polarizability of the proton beam will permit 
investigation of helicity flip at the nucleon vertex~\cite{hep99_07_568}\hspace*{-2.2ex}.\hspace*{2.2ex} The
H1 and ZEUS studies have no access to such information, which is essential
for the complete understanding of the helicity structure of this diffractive
interaction, especially for the semi-exclusive processes described below.

\section{Semi-exclusive processes}
\label{sec:semi-exclusive}
The photoproduction of light vector mesons with high transverse momenta
has raised interest recently~\cite{pl_375_301,pr_53_3564,*pr_54_5523,pl_449_306} as a means of investigating the r\^ole of 
the momentum transfer $t$ in establishing the scale of the interaction,
since the photoproduction of {\rhoz} mesons at low {\abst} has been
shown to be governed by a soft diffractive production mechanism (see
Sect.~\ref{sec:exphotoproduction}). However, diffractive photoproduction
of the light vector 
mesons has been shown to be dominated by the proton-dissociative process
for momentum transfers ${\abst}\gsim 0.5~\gevsq$\hspace*{-1.5ex},\hspace*{1.2ex} well
below the perturbative regime.\hspace*{-0.7ex}~\cite{ichep98_788} The investigation of exclusive vector-meson
production at high {\abst} requires a trigger on the elastically
scattered proton and such studies at HERA have lacked the integrated
luminosity necessary to measure the small cross sections at such high
values of {\abst}.\hspace*{-0.7ex}~\footnote{Clearly, a central detector together with a forward
proton spectrometer at the high-luminosity EPIC collider would 
provide such measurements.} However, the ZEUS collaboration~\cite{eps99_499} has recently
employed a photoproduction-tagging method to investigate the
proton-dissociative production of {\rhoz} mesons for values of {\abst} up to
11~{\gevsq}\hspace*{-1.2ex}.\hspace*{1.2ex} The trigger conditions
required the scattered positron to be detected in a special-purpose
tungsten/scintillator calorimeter located 3~cm from the positron beam axis, 44~meters distant
from the nominal e$^+$p interaction point in the positron-beam flight direction.
The position of this photoproduction tagger determines the accepted range
of energy
lost by the positron to the photon which interacts with the proton, thus
restricting the $W$ range to the region $80<W<120~\gev$.
Since the transverse momentum of the final-state positron is thus required to
be small (${\qsq}<0.01~{\gevsq}$), the transverse momentum of the
{\rhoz} ($p_{\rm t}$) detected in the central detector via its dipion
decay provides an accurate approximation
for the square of the momentum transferred to the proton ($t$): $t\simeq
-p^2_{\rm t}$. 
Offline data selection criteria include the reconstruction of exactly two
tracks from the interaction vertex and reject events with calorimetric energy
deposits in the rear and barrel sections of the calorimeter which are not
associated with the extrapolation of either track. The selected events exhibit
a semi-exclusive topology with a substantial rapidity
gap between the dissociated nucleonic 
system and the two tracks. Even
for the highest values of {\abst}, the decay pions are in the rear half of the central detector.

\subsection{Differential cross sections}
\begin{figure}[ht]
\begin{center}
\epsfig{file=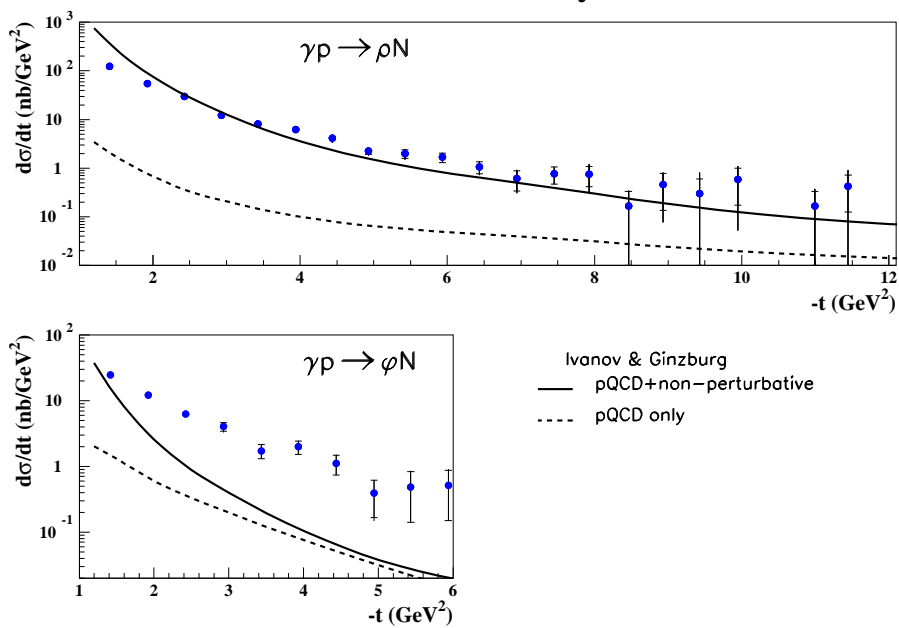, width=\textwidth, bbllx=18, bblly=228,
  bburx=540, bbury=618, clip=}
\caption{
\label{fig:rho}
Preliminary measurements
of the differential cross sections $\frac{\D\sigma}{\D t}$ for
the proton-dissociative photoproduction of {\rhoz} and $\phi$ mesons in the energy range \mbox{$80 < W
  <120~\gev$} recently presented by the ZEUS collaboration~\protect\cite{eps99_499}\hspace*{-2.2ex}.\hspace*{2.2ex} The inner error bars represent the statistical uncertainties;
the outer error bars show the quadratic sum of statistical and systematic uncertainties. 
The results of a QCD calculation by Ivanov and
Ginzburg~\protect\cite{pr_53_3564,*pr_54_5523} (solid line) which
distinguishes the perturbative contribution (dashed line) are shown for
comparison \vspace*{-5mm}
}
\end{center}
\end{figure}

Differential cross sections $\frac{\D\sigma}{\D t}$ were obtained  for
{\rhoz}, $\phi$ and  {\jpsi} mesons. Figure~\ref{fig:rho} shows the 
cross sections measured for
{\rhoz} and $\phi$ mesons. These exhibit an impressively hard spectrum;
a fit to the form $(-t)^{-n}$
results in a value for $n$ of approximately~3.
The results are compared to a QCD calculation by Ivanov
and Ginzburg~\cite{pr_53_3564,*pr_54_5523} which estimates both perturbative
and nonperturbative contributions. In this model, the nonperturbative
contribution is found account for the hardness of the spectrum and
to dominate in the region covered by the measurements
for the {\rhoz} meson. 
The calculation underestimates the $\phi$ cross section over most of the $t$ range covered. 
The comparison of the data to the calculation of the
perturbative contribution alone shows a good description of the shape of the $t$
dependence, whereas the magnitude of the cross section is underestimated by more
than one order of magnitude. 
The model thus fails to reproduce the ratio of $\phi$ production to that for the {\rhoz}
meson at high {\abst}, which is measured to be consistent with
the value of 2/9 expected according to 
the assumption of a direct point-like coupling
of the photon to the quark constituents of the vector mesons.\hspace*{-0.7ex}~\cite{eps99_499} These
measurements thus permit sensitive tests of perturbative models and provide
information on the internal structure of the vector mesons.
\subsection{Helicity analyses}
\label{sec:semihelicity}
The ZEUS collaboration has employed this 
photoproduction-tagging technique to
perform decay-angle
analyses of  diffractive
photoproduction~\cite{hep99_06_005,eps99_499} of pion pairs in the {\rhoz}
mass
region at values of {\abst} up to 
4~{\gevsq}\hspace*{-1.2ex}.\hspace*{1.2ex} In this study, the small contribution from the elastic process was
not subtracted. The decay-angle distribution was parametrized in terms of combinations
of spin-density matrix elements in the Schilling-Wolf convention~\cite{np_61_381}\hspace*{-2.2ex},\hspace*{2.2ex}
$r^{04}_{ij}$, as
\begin{eqnarray}
\begin{array}{ll}{W(\thetah,\phih)}={3\over 4\pi} \biggl[
\frac{1}{2} \left(1-{{r^{04}_{00}}}\right)+\frac{1}{2} 
\left(3{{r^{04}_{00}}}-1\right)\cos^2{\thetah} \\[4mm]
-\sqrt{2}{{{\rm Re}(r^{04}_{10})}} \sin 2\thetah\cos\phih
-{{r^{04}_{1-1}}}\sin^2\thetah\cos{2\phih}
\;\biggr],\end{array}
\end{eqnarray}
where the three-dimensional distribution (Eq.~\ref{full_equation}) has been
averaged over the unmeasured azimuthal angle between the positron scattering
plane and the {\rhoz} production plane, and thus no longer distinguishes the
photon helicity states~$\pm 1$.

Under the assumption that the dipion final state is produced with one
unit of angular momentum, and neglecting the contribution by longitudinal
photons, these combinations of matrix elements are
related to the helicity amplitudes, \mbox{$T_{\lambda_{\rho} \lambda_{\gamma}}$}, as follows~\cite{np_61_381,dr_99_102}:
\begin{eqnarray}
r_{00}^{04} \simeq \frac{\strut T_{01}^2}{\strut
    T_{01}^2+T_{11}^2+T_{1-1}^2},\hspace*{5mm}
r_{1-1}^{04} \simeq \frac{\strut {\rm Re}(T_{11}T_{1-1}^*)}{\strut
    T_{01}^2+T_{11}^2+T_{1-1}^2}, \nonumber \\[3mm]
{\rm Re}(r_{10}^{04}) \simeq \frac{1}{2} \left [ \frac
{\strut {\rm Re}(T_{11}T_{01}^*) + {\rm Re}(T_{1-1}T_{0-1}^*)}{\strut T_{01}^2+T_{11}^2+T_{1-1}^2} 
\right].
\end{eqnarray}

\begin{figure}[htbp]
\begin{center}
\epsfig{file=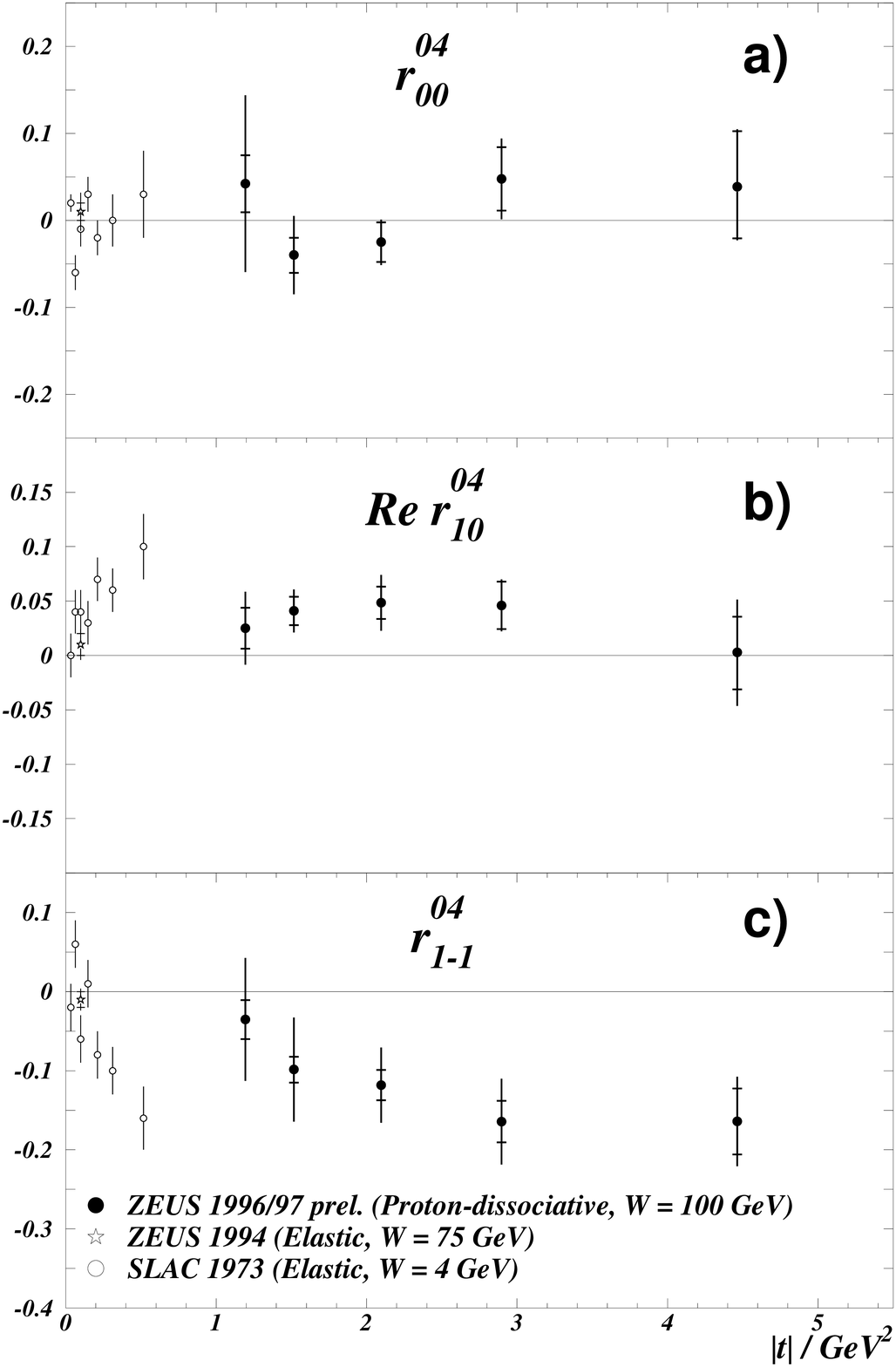, width=0.8\textwidth, bbllx=0, bblly=0, bburx=800, bbury=1227, clip=}
\caption{
Recent measurements by the ZEUS  
collaboration~\protect\cite{hep99_06_005,eps99_499} 
of the combinations of matrix elements  (a)~$r^{04}_{00}$, (b)~Re~$r^{04}_{10}$, (c)~$r^{04}_{1-1}$ for the
  diffractive  photoproduction of pion pairs. 
The inner error bars represent the statistical uncertainties;
the outer error bars show the statistical and systematic uncertainties added
in quadrature. See text for full description}
\label{fig:rij}
\end{center}
\end{figure}

Figure~\ref{fig:rij}
shows the results for the combinations
of matrix elements obtained from a least-squares minimization procedure
in which they served as fit parameters.
The inner error bars represent the statistical
uncertainty; the outer error bars show the quadratic sum of
statistical and systematic uncertainties. 
The systematic
uncertainties are dominated by the uncertainty in the acceptance corrections.
The dipion mass range was restricted to the region
\mbox{$0.45\hspace*{-2pt}<\hspace*{-2pt}M_{\pi\pi}\hspace*{-2pt}<\hspace*{-2pt}1.1~{\gev}$}.
The results are compared to the results at lower {\abst} for the exclusive
reaction obtained with 9 {\gev}
photons from a backscattered laser beam  incident on a hydrogen bubble 
chamber at SLAC.\hspace*{-0.7ex}~\cite{pr_7_3150} Also shown are the ZEUS 1994 results for exclusive {\rhoz}
photoproduction at low {\abst}.\hspace*{-0.7ex}~\cite{epj_2_247} 
The parameter $r^{04}_{00}$
is consistent with zero over the entire range in {\abst}. 
The combination \mbox{Re $r^{04}_{10}$}, which is predominantly sensitive to the interference between
the helicity-conserving amplitude and the single-flip amplitude, shows slight
evidence for a single-flip contribution in both the SLAC data and the
high-{\abst} ZEUS results. A clear indication of a double-flip contribution is
shown by the measurements of $r^{04}_{1-1}$ at high {\abst}, as was seen in
the SLAC results at lower {\abst}. 

In order to estimate the effect on the angular distributions of a 
hypothesized dipion background to {\rhoz} decay, the decay-angle 
analysis was repeated for restricted dipion mass ranges above
(\mbox{$0.77<M_{\pi\pi}<1.0~{\gev}$}) and below
(\mbox{$0.6<M_{\pi\pi}<0.77~{\gev}$}) the nominal 
value for the {\rhoz} mass. 
The observed value for each of the combinations of 
matrix elements was found to depend
significantly
on the mass range chosen, though the above conclusions concerning the lack of
longitudinal polarization and the evidence for a double-flip contribution
remain unchanged. This
dependence on dipion invariant mass may suggest the presence of a nonresonant background. The extraction of the spin-density
matrix elements for the {\rhoz} meson alone 
from these dipion angular distributions thus
awaits the understanding of this dependence.

The complexity of this helicity structure, which the SLAC results showed
to hold also at an energy lower than that of EPIC, 
is thus shown to persist at high values of {\abst}
where the hardness of the {\abst} spectrum and the $\phi$/{\rhoz} ratio
encourage attempts to describe this semi-exclusive process in the 
framework of perturbative QCD. The physics program 
at the EPIC collider, which emphasizes
high luminosity and polarizability over high energy, will permit
the conclusive resolution of these  fundamental quandaries
in the understanding of the strong interaction. 
\section{Concluding remarks}
This report has presented a selection of investigations
by the H1 and ZEUS collaborations into diffractive
processes in electron-proton interactions, using results obtained during
the first six years of operation of the HERA collider. These studies have 
provided a great deal of information pertaining to the applicability
of quantum chromodynamical descriptions of diffraction. Particular emphasis
has been placed on relevance to the physics program at the proposed EPIC 
electron/polarized-ion collider, bringing investigations which require
a wide energy range within the asymptotic region at high energy, and those of the
helicity structure of diffraction to the fore. The rich phenomenology of
examples of inclusive, exclusive and semi-exclusive processes in
electron-proton interactions has been discussed.
However, this synopsis
cannot pretend to have been an exhaustive list of the open questions
which are to be addressed by experiments at the EPIC collider, not even of
those which have already begun to be investigated at HERA. One need merely
consider the issues of the dependence of the exponential slopes in $t$
at low {\abst} on energy and photon virtuality in both exclusive and
semi-exclusive processes, of vector-meson production ratios, of
photo- and electroproduction of radially excited quarkonia and of production
processes involving the exchange of quantum numbers other than those of the
vacuum to recognize the incompleteness of the present summary. The
experimental program of the collider experiments at HERA has made obvious
the power of such investigations to elucidate the 
properties of the strong interaction. Experimental investigations 
at the EPIC collider are certain to provide high statistical accuracy 
in the kinematic region which must be further investigated 
in order to make progress in our understanding of strong-interaction
dynamics.

\section*{Acknowledgments}
The author gratefully acknowledges support from the workshop organizers. This
work is also supported by the Federal Ministry of Education and Research of
Germany.
\vfill

\bibliographystyle{zeusstylem}
\bibliography{zeuspubs,h1pubs,otherpubs}

\end{document}